\documentclass{article}

\usepackage{amsmath}
\usepackage{amssymb}
\usepackage{amsthm}
\usepackage{calc}
\usepackage{url}

\usepackage[margin=2.9cm,a4paper]{geometry}

\newtheorem{theorem}{Theorem}
\newtheorem{lemma}[theorem]{Lemma}

\newtheorem{definition}[theorem]{Definition}

\newtheorem{remark}[theorem]{Remark}

\theoremstyle{plain}

\newcommand*{\cA}{\mathcal{A}}
\newcommand*{\cD}{\mathcal{D}}
\newcommand*{\cP}{\mathcal{P}}
\newcommand*{\cS}{\mathcal{S}}
\newcommand*{\cDp}{\mathcal{D'}}
\newcommand*{\eps}{\varepsilon}

\newcommand*{\real}{\mathrm{real}}
\newcommand*{\ideal}{\mathrm{ideal}}

\newcommand*{\Iacc}{I_{\mathrm{acc}}}

\newcommand*{\ket}[1]{| #1 \rangle}
\newcommand*{\bra}[1]{\langle #1 |}

\newcommand*{\proj}[1]{\ket{#1}\bra{#1}}

\newcommand*{\system}[5]{\linethickness{1pt}
                        \put(#1,#2){\framebox(#3,#4)[cc]{#5}}}
\newcommand*{\systemd}[5]{\linethickness{1pt}
                        \put(#1,#2){\dashbox{1.5}(#3,#4)[cc]{#5}}}
\newcommand*{\pr}[3]{\put(#1,#2){\makebox(0,0){#3}}}
\newcommand*{\prl}[3]{\put(#1,#2){\makebox(0,0)[l]{#3}}}

\newcommand*{\arrowwidth}{\linethickness{1pt}}
\newcommand*{\arrowupp}{\makebox{\vphantom{X}}{\hspace{-0.7ex}$\blacktriangle$}}
\newcommand*{\arrowdownp}{\vspace{0.28ex}\hspace{-0.7ex}$\blacktriangledown$}
\newcommand*{\arrowrightp}{\vspace{0ex}\hspace{-0.5ex}$\blacktriangleright$}

\newcounter{tempx}\newcounter{tempy}

\newcommand*{\arrowd}[3]{\arrowwidth
                        \setcounter{tempx}{#3-5}
                        \setcounter{tempy}{#2+5}
                        \put(#1,\value{tempy}){\line(0,1){\value{tempx}}}
                        \put(#1,#2){\makebox(0,0)[b]{ \arrowdownp   }}
}
\newcommand*{\arrowr}[3]{\arrowwidth
                        \setcounter{tempx}{#3-5}
                        \put(#1,#2){\line(1,0){\value{tempx}}}            
                        \put(#1,#2){\makebox(#3,0)[r]{ \arrowrightp  }}
}

\newcommand*{\arrowv}[3]{\arrowwidth
                        \setcounter{tempx}{#3-10}
                        \setcounter{tempy}{#2+5}          
                        \put(#1,\value{tempy}){\line(0,1){\value{tempx}}}
                        \put(#1,#2){\makebox(0,#3)[t]{
                          \arrowupp  }}          
                        \put(#1,#2){\makebox(0,0)[b]{
                          \arrowdownp   }}
}

\title{Composability in Quantum Cryptography}

\author{J\"orn M\"uller-Quade and Renato Renner}

\date{}

\begin{document}

\maketitle

\begin{abstract}

  If we combine two secure cryptographic systems, is the resulting
  system still secure? Answering this question is highly non-trivial
  and has recently sparked a considerable research effort, in
  particular in the area of classical cryptography. A central insight
  was that the answer to the question is yes, but only within a well
  specified \emph{composability framework} and for carefully chosen
  security definitions.

  In this article, we review several aspects of composability in the
  context of \emph{quantum} cryptography.  The first part is devoted
  to key distribution. We discuss the security criteria that a quantum
  key distribution protocol must fulfill to allow its safe use within
  a larger security application (e.g., for secure message
  transmission); and we demonstrate|by an explicit example|what can go
  wrong if conventional (non-composable) security definitions are
  used. Finally, to illustrate the practical use of composability, we
  show how to generate a continuous key stream by sequentially
  composing rounds of a quantum key distribution protocol.

  In a second part, we take a more general point of view, which is
  necessary for the study of cryptographic situations involving, for
  example, mutually distrustful parties. We explain the universal
  composability framework and state the composition theorem which
  guarantees that secure protocols can securely be composed to larger
  applications. A focus is set on the secure composition of quantum
  protocols into unconditionally secure classical protocols.  However,
  the resulting security definition is so strict that some tasks
  become impossible without additional security assumptions. Quantum
  bit commitment is impossible in the universal composability
  framework even with mere computational security. Similar problems
  arise in the quantum bounded storage model and we observe a
  trade-off between the universal composability and the use of the
  weakest possible security assumptions.

\end{abstract}
  
\section{Introduction}

Provable security, even for complex security applications, is
desirable. However, giving one monolithic security proof for a larger
cryptosystem is error prone, and a modular design is usually
advantageous. But this comes with a major difficulty, namely that
security definitions are not generally closed under
composition. Therefore, an application may be insecure even if the
individual components it consists of are secure. During the past few
years, finding solutions to this problem has been a main focus of
research in cryptography. This research effort has resulted in the
development of frameworks in which security definitions are
\emph{universally composable}.

We review several aspects of composability in the context of quantum
cryptography and structure our exposition into two
parts. Section~\ref{sec:QKD} considers the security and composability
of Quantum Key Distribution (QKD), which is the most prominent
application of quantum cryptography. In a second part, starting with
Section~\ref{sec:general}, we consider the problem of composability
for general security applications.

The reason for this organization of the paper is that for the usual
treatment of QKD, one assumes a fixed \emph{adversary structure},
i.e., Alice and Bob are always honest (in particular, they trust each
other), while only a third party with access to the communication
channels is malicious. This avoids many of the problems that arise in
the more general considerations outlined in Sections~\ref{sec:general}
through~\ref{sec:uc}, where arbitrary parties may be corrupted.

\section{Quantum Key Distribution (QKD)} \label{sec:QKD}

\subsection{QKD in a Nutshell}

\emph{Quantum key distribution (QKD)} is the art of distributing a
secret key to two distant parties, \emph{Alice} and \emph{Bob},
connected by an insecure quantum channel. Technically, a \emph{secret
  key} is simply a random bitstring for which there is a certain
guarantee that its value is unknown to an adversary, \emph{Eve}. Such
a key may be used for a variety of cryptographic tasks. The most
prominent among them is certainly the secure transmission of secret
messages over an insecure channel. Here, the key typically serves as a
one-time-pad for message encryption.

In the past two decades, numerous QKD schemes have been
proposed. Although they differ in many aspects (such as their
realizability with current technology), they still very much resemble
the original protocols put forward by Bennett and
Brassard~\cite{BenBra84} (based on ideas by Wiesner~\cite{Wiesner83})
and by Ekert~\cite{Ekert91}. We will not attempt here to give a
description of these protocols. In fact, for the purpose of this
article, it is sufficient to take a rather abstract point of view,
where the internal workings of the protocols are unimportant. (The
reader interested in the concrete protocols is referred to the
original articles~\cite{BenBra84,Ekert91} as well as the recent review
articles~\cite{SBCDLP08} and references therein.)

The security of QKD basically relies on an intrinsic property of
quantum mechanics, namely that it is generally impossible to copy the
state of a system without disturbing the original.\footnote{More
  precisely, it is impossible to build a physical device that takes as
  input an unknown quantum state and outputs two copies of it. This
  impossibility is also known as \emph{non-cloning theorem}. For QKD,
  it is important to have a quantitative version of this statement,
  sometimes called \emph{information-disturbance trade-off.}} For
cryptography, this means that any attempt of an attacker to ``steal''
appropriately encoded information can in principle be detected. This
also motivates the basic structure of QKD protocols: first, Alice and
Bob send random signals over the quantum channel and then, in a second
step, perform tests to check for disturbances in the signals, which
may be a sign of an attack. Depending on this test, the protocol
typically has one of two different outcomes. Either the disturbances
are found to be too large, in which case the protocol aborts with the
declaration that no key can be generated. Otherwise, if there are no
(or only small) disturbances, Alice and Bob use the randomness in the
distributed signals to generate a key.\footnote{More generally, a
  protocol may generate keys whose length depends on an estimate of
  the maximum amount of information that an adversary may have gained
  by an eavesdropping attack.}

Although QKD is often said to be \emph{unconditionally secure}, there
are still a few assumptions needed to prove security of the generated
keys. The first (usually implicit in the literature) is that Alice and
Bob are \emph{honest}, meaning that they both follow their respective
part of the protocol.\footnote{Dropping this assumption leads to the
  additional problem of generating randomness by mutually mistrustful
  parties, which is known as \emph{coin flipping}~\cite{Blum83}.}
Second, it is assumed that Alice and Bob can exchange classical
messages \emph{authentically}, i.e., it is impossible for an adversary
to alter the classical messages exchanged between Alice and Bob. In
practice, this is usually achieved by invoking an authentication
scheme~(see, e.g., \cite{WegCar81}) which, however, requires Alice and
Bob to share a short initial key. Because of this latter assumption,
QKD is sometimes called key \emph{growing} rather than key
\emph{distribution}.

After this brief introduction, we are now ready to have a closer look
at the notion of security used in the context of QKD. We introduce an
explicit definition (Section~\ref{sec:security}) and then show its
composability (Section~\ref{sec:comp}). As an example, we discuss the
problem of generating a continuous key stream by sequentially
composing many rounds of a QKD protocol
(Section~\ref{sec:keystream}). We then conclude the part on QKD with
an example that pinpoints the problems arising when employing a
non-composable security definition, which incidentally has been widely
used in the literature (Section~\ref{sec:attack}).

\subsection{Security Criteria} \label{sec:security}

To define security, we first need to have a clearer picture of what a
QKD protocol is supposed to do. We start with a list of the properties
we expect an \emph{ideal} protocol to have and then, in a second step,
define security of \emph{real} protocols by their indistinguishability
from the ideal case. In accordance with the terminology used in the
context of multi-party computation, we call these properties
\emph{secrecy}, \emph{correctness}, and \emph{robustness} (see
also~\cite{Kao08}). We denote by $S_A$ and $S_B$ the final outputs of
the protocol on Alice and Bob's side, respectively. Following the
discussion above, the protocol may either generate keys, in which case
$S_A$ and $S_B$ are two identical random bitstrings of a certain fixed
length $\ell$, or it may abort, in which case we set $S_A = \perp$ and
$S_B = \perp$.\footnote{Alternatively, the length $\ell$ of the
  generated key may be determined during the run of the protocol, with
  $\ell = 0$ if the protocol aborts (see, e.g., \cite{BHLMO05}). For
  practical applications, however, it is usually more convenient to
  work with a fixed key length.}  Furthermore, we denote by $E$ the
entire (quantum) system controlled by an adversary. In particular, $E$
contains all the information that the adversary acquires during the
run of the protocol.

We consider here the strongest type of security, namely \emph{security
  against general attacks}. This means that an adversary may
arbitrarily tamper with the signals exchanged between Alice and Bob
over the quantum channel.\footnote{One sometimes restricts the
  security analysis to more restricted types of attacks. An example
  are \emph{collective attacks}~\cite{BihMor97a}, where it is assumed
  that the adversary acts on each of the signals sent through the
  channel independently and identically. This is useful because, for
  most protocols, security against collective attacks implies security
  against general attacks~\cite{Renner05,Renner07}.} In addition, she
may eavesdrop (but not alter) the classical communication. We also
introduce the notion of a \emph{passive} adversary, who does not
disturb the quantum communication. Formally, this simply means that
the behavior of the quantum channel is described by a fixed noise
model. For QKD based on qubit-systems, for instance, the standard is
to consider channels that introduce random bit- and phase-flips (with
a given probability).

\paragraph{Perfect Security.}

We now say that a QKD scheme is \emph{perfectly secure} if the
following holds for any attack.

\bigskip

\noindent \emph{Correctness:} The outputs of the protocol on Alice and Bob's
side are identical (i.e., $S_A = S_B$).

\bigskip

\noindent \emph{Secrecy:} If the protocol produces a key $S_A$ (i.e., if
$S_A \neq \perp$) then $S_A$ is uniformly distributed and independent
of the state of the system $E$ held by the adversary.\footnote{Because
  of the \emph{correctness property}, it is sufficient to require
  secrecy for either $S_A$ or $S_B$.}

\bigskip

\noindent \emph{Robustness:} If the adversary is passive then a key is
generated (i.e., $S_A \neq \perp$).\footnote{Note that this property
  is always relative to a given noise model of the quantum channel.}

\bigskip

It is easy to see that none of these criteria can be dropped without
making the task trivial. In fact, without the correctness requirement,
a protocol may just produce uncorrelated randomness on Alice and Bob's
side. Similarly, without the robustness requirement, a protocol may
always output $S_A = S_B = \perp$.

\paragraph{Approximate Security.}

Unfortunately, it is (provably) impossible to design a QKD protocol
that is perfectly secure according to the above definition. One thus
typically considers a relaxation where the requirement is that the
behavior of the scheme is \emph{similar} (but not necessarily
\emph{equal}) to an idealized scheme which is perfectly secure. This
can be made precise using the notion of \emph{indistinguishability}.

More specifically, one considers a hypothetical device, called
\emph{distinguisher}, which interacts with either the real protocol,
in the following denoted $\cP^{\real}$, or an ideal protocol,
$\cP^{\ideal}$, and then outputs a \emph{guess bit} $B$. The
distinguisher may have access to all regular inputs and outputs of the
protocol (in our case, we only have outputs, namely $S_A$ and $S_B$)
as well as to the system $E$ normally controlled by the adversary. We
say that $\cP^{\real}$ and $\cP^{\ideal}$ are
\emph{$\eps$-indistinguishable} for $\eps \geq 0$ if, for any such
distinguisher,
\begin{align} \label{eq:distadv}
  \Pr[B=1 | \cP^{\real}] - \Pr[B=1 | \cP^{\ideal}] \leq \eps \ .
\end{align}
Here $\Pr[B=1 | \cP^{\real}]$ and $\Pr[B=1 | \cP^{\ideal}]$ denote the
probabilities that the distinguisher's output $B$ equals $1$ when
interacting with $\cP^{\real}$ and $\cP^{\ideal}$, respectively.

The notion of $\eps$-indistinguishability naturally leads to the
following definition of $\eps$-security.

\begin{definition} \label{def:security} A QKD protocol $\cP^{\real}$
  is \emph{$\eps$-secure} if it is $\eps$-indistinguishable from a
  (hypothetical) protocol $\cP^{\ideal}$ which is perfectly secure,
  i.e., $\cP^{\ideal}$ satisfies the correctness, the secrecy, and the
  robustness criteria above.
\end{definition}

Intuitively, the parameter $\eps$ can be understood as the
\emph{maximum failure probability} of the protocol $\cP^{\real}$,
i.e., the maximum probability that $\cP^{\real}$ deviates from the
behavior of the ideal protocol $\cP^{\ideal}$.\footnote{This intuition
  can be made precise in a purely classical context~\cite{Maurer02}.}
For practical considerations, is often useful to quantify the
correctness, secrecy, and robustness of a protocol separately. The
following definition is an obvious generalization of the above.

\begin{definition} \label{def:securityq} A QKD protocol is
  \emph{$\eps$-correct}, \emph{$\eps$-secret}, or \emph{$\eps$-robust}
  if it is $\eps$-indistinguishable from a perfectly correct, secure,
  or robust scheme, respectively.
\end{definition}

\begin{remark}
One can show that, if a protocol is $\eps_c$-correct, $\eps_s$-secret,
and $\eps_r$-robust then it is $\eps$-secure, for $\eps = \eps_c +
\eps_s + \eps_r$.
\end{remark}

The requirements on the different parameters are generally quite
diverse.  Typically, a relatively large value $\eps_r$ for the
robustness (e.g., $\eps_r = 0.1$) can be tolerated, because the
protocol may just be repeated in case it does not generate a key. In
contrast, the parameter $\eps_s$ for the secrecy can be interpreted as
the (maximum) probability by which an adversary may get secret
information without being detected, which one typically wants to keep
small (e.g., $\eps_s = 10^{-10}$).

It is easy to see that $\eps$-correctness is equivalent to the
requirement that the outputs $S_A$ and $S_B$ produced by the protocol
on Alice and Bob's side differ only with small probability,
\begin{align} \label{eq:correctness}
  \Pr[S_A \neq S_B] \leq \eps \ .
\end{align}
Similarly, for $\eps$-robustness, the requirement is that
\begin{align} \label{eq:robustness}
  \Pr[S_A = \perp] \leq \eps
\end{align}
holds whenever the adversary is passive. The situation is a bit more
subtle (and more interesting) for the secrecy criterion, which can be
made more concrete as follows.

Let $\cS := \{0,1\}^\ell$ be the key space, i.e., the output $S_A$
takes values in the set $\cS \cup \{\perp\}$. Furthermore, for any
fixed value $s \in \cS \cup \{\perp\}$ of $S_A$, let the state of the
system $E$ be denoted by $\rho_E^s$. The joint state of $S_A$ and $E$
can then be represented as a cq-state\footnote{The state of a
  bipartite system is called \emph{classical-quantum (cq)} if the
  first subsystem is purely classical (in the sense that its states
  are perfectly distinguishable.)}
\begin{align*}
  \rho_{S_A E} = \sum_{s \in \cS \cup \{\perp\}} p_s \proj{s} \otimes \rho_E^s
\end{align*}
where $p_s$ is the probability that $S_A = s$ and where
$\{\ket{s}\}_{s \in \cS \cup \{\perp\}}$ is a family of orthonormal
vectors. It is easy to see that, for any attack, the state resulting
from the run of a perfectly secure scheme has the form
\begin{align} \label{eq:idealstate} \rho_{S_A E}^{\mathrm{perfect}} =
  (1-p_\perp) \sum_{s \in \cS} \frac{1}{|\cS|} \proj{s} \otimes \rho'_E
  + p_\perp \proj{\perp} \otimes \rho''_E \ ,
\end{align}
where $p_\perp \in [0,1]$ and where $\rho'_E$ and $\rho''_E$ are
density operators. With these definitions, we arrive at a
reformulation of $\eps$-secrecy in terms of the trace
distance~\cite{RenKoe05,BHLMO05}.\footnote{Lemma~\ref{lem:secrecy} is
  an immediate consequence of the well known one-to-one relation
  between the indistinguishability of two quantum states and their
  trace distance.}

\begin{lemma} \label{lem:secrecy}
  A QKD protocol is $\eps$-secret if and only if, for any attack, the
  cq-state $\rho_{S_A E}$ describing the joint state of the protocol
  output $S_A$ and the system $E$ held by the adversary satisfies
  \begin{align} \label{eq:seccrit}
    \frac{1}{2} \bigl\| \rho_{S_A E} - \rho_{S_A E}^{\mathrm{perfect}} \bigr\|_1 \leq \eps
  \end{align}
  for some state $\rho_{S_A E}^{\mathrm{perfect}}$ of the form~\eqref{eq:idealstate}.
\end{lemma}

In security proofs, correctness and secrecy are usually established by
separate arguments. While the correctness parameter $\eps_c$ is
essentially determined by the quality of the \emph{error correction}
procedure used to reconcile the raw keys, the secrecy $\eps_s$ rests
upon various other elements of the protocol. In the simplest case,
$\eps_s$ is a function of the accuracy of the \emph{estimation
  procedure}, which measures the disturbances of the transmitted
signals, as well as of the parameters of the \emph{privacy
  amplification step}, which is used to transform the (partially
secret) raw key into a final secret key satisfying~\eqref{eq:seccrit}.

\subsection{Composing QKD with Other Cryptographic
  Primitives} \label{sec:comp}

Since a secret random string is of little interest by itself, QKD is
almost never used as a stand-alone application. Instead, one typically
is interested in higher cryptographic tasks such as secure message
transmission. QKD then just serves as a mechanism to provide the key
material needed by the application. In addition, QKD often is built on
top of other cryptographic primitives such as authentication schemes,
whose task is to make sure the adversary cannot alter the classical
messages sent over the insecure channel. Hence, composability of the
underlying security definitions is vital in the context of QKD.

\paragraph{What Does Composability Mean?}

To get a more precise understanding of the notion of composability in
the context of QKD, we consider a situation where the key produced by
a QKD protocol $\cP^{\real}$ is later used in an application
$\cA^{\real}$, e.g., an encryption scheme. Assume that the protocol
$\cP^{\real}$ is $\eps_1$-secure, and let the application
$\cA^{\real}$ be $\eps_2$-secure, i.e., $\eps_2$-indistinguishable
from an idealized application $\cA^{\ideal}$. The claim then is that
the composite system, denoted $\cA^{\real} \circ \cP^{\real}$, where
the application $\cA^{\real}$ is fed with the key produced by
$\cP^{\real}$, is $\eps$-secure, for $\eps = \eps_1 + \eps_2$.

The claim becomes even simpler in the special case where $\cA^{\real}$
is based on one-time-pad encryption. When being fed with a perfectly
secret key, one-time-pad encryption is indistinguishable from a
perfect encryption procedure, which simply produces a ciphertext that
is statistically independent of the message. We thus have $\eps_2 =
0$. Hence, according to the above claim, when one-time-pad encryption
is combined with an $\eps_1$-secure QKD protocol $\cP^{\real}$, the
resulting scheme is $\eps_1$-secure. That is, it produces ciphertexts
which are $\eps_1$-indistinguishable from uniform randomness.

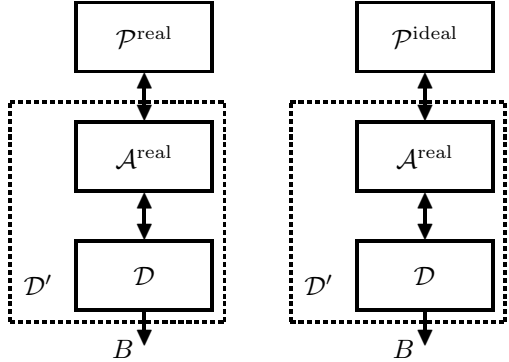
\begin{figure}
\begin{center}
\begin{picture}(190,130)(0, -10)

  \system{25}{95}{50}{25}{$\cP^{\real}$}
  \system{130}{95}{50}{25}{$\cP^{\ideal}$}
  
  \arrowv{50}{75}{20}
  \arrowv{155}{75}{20}

  \system{25}{50}{50}{25}{$\cA^{\real}$}
  \system{130}{50}{50}{25}{$\cA^{\real}$}

  \arrowv{50}{30}{20}
  \arrowv{155}{30}{20}

  \system{25}{5}{50}{25}{$\cD$}
  \system{130}{5}{50}{25}{$\cD$}

  \systemd{0}{0}{80}{83}{}
  \prl{5}{14}{$\cDp$}
  \systemd{105}{0}{80}{83}{}
  \prl{110}{14}{$\cDp$}
 
  \arrowd{50}{-10}{15}
  \prl{38}{-10}{$B$}
  \arrowd{155}{-10}{15}
  \prl{143}{-10}{$B$}
  
\end{picture}
\end{center}

\caption{\label{fig:distinguish} {\bf Indistinguishability.} The
  combination of the original distinguisher~$\cD$ with $\cA^{\real}$
  gives a new distinguisher~$\cDp$ for~$\cP^{\real}$
  and~$\cP^{\ideal}$.}
\end{figure}

\paragraph{Why Is Our Definition Composable?} Roughly speaking, the
security parameters $\eps_1$ and $\eps_2$ can be understood as the
maximum failure probabilities of $\cP^{\real}$ and $\cA^{\real}$,
respectively (see the paragraph after
Definition~\ref{def:security}). Hence, according to the union bound,
if one combines $\cP^{\real}$ and $\cA^{\real}$, the total failure
probability cannot be larger than $\eps = \eps_1 + \eps_2$. This
already gives an intuitive understanding why the combined scheme
$\cA^{\real} \circ \cP^{\real}$ is $\eps$-secure, as claimed above.

We will now give a slightly more rigorous argument for this
claim. Assume by contradiction that the composite system $\cA^{\real}
\circ \cP^{\real}$ is not $\eps$-indistinguishable from $\cA^{\ideal}
\circ \cP^{\ideal}$, i.e., there exists a distinguisher $\cD$ whose
output $B$ satisfies
\begin{align} \label{eq:distcomb}
  \Pr[B=1|\cA^{\real} \circ \cP^{\real}] - \Pr[B=1|\cA^{\ideal} \circ \cP^{\ideal}]
> 
  \eps = \eps_1 + \eps_2
\end{align}
(cf.~\eqref{eq:distadv}). Assume now that we use the same
distinguisher $\cD$ to distinguish $\cA^{\real} \circ \cP^{\real}$
from $\cA^{\real} \circ \cP^{\ideal}$, where the latter denotes the
composite scheme consisting of the real application fed with a key
produced by a perfect QKD scheme. Because $\cA^{\real}$ is identical
in both cases, we can alternatively treat $\cA^{\real}$ as part of a
(more complex) distinguisher $\cDp$ which now interacts with either
$\cP^{\real}$ or $\cP^{\ideal}$ (see
Fig.~\ref{fig:distinguish}). Because, by assumption, $\cP^{\real}$ is
$\eps_1$-secure and, hence, $\eps_1$-indistinguishable from
$\cP^{\ideal}$, we find
\begin{align} \label{eq:distone}
  \Pr[B=1|\cA^{\real} \circ \cP^{\real}] - \Pr[B=1|\cA^{\real} \circ \cP^{\ideal}]
\leq
  \eps_1 \ .
\end{align}
Similarly, because $\cA^{\real}$ is $\eps$-indistinguishable from
$\cA^{\ideal}$, we find
\begin{align} \label{eq:disttwo}
  \Pr[B=1|\cA^{\real} \circ \cP^{\ideal}] - \Pr[B=1|\cA^{\ideal} \circ \cP^{\ideal}]
\leq
  \eps_2 \ .
\end{align}
Combining~\eqref{eq:distone} and~\eqref{eq:disttwo}
contradicts~\eqref{eq:distcomb} and, hence, concludes our proof of
composability.

\subsection{Example Application: Generating a Continuous Key
  Stream} \label{sec:keystream}

As already mentioned, composability of the keys produced by a QKD
scheme is crucial because these are typically used in further
applications. Here, we consider their use for authentication in
subsequent rounds of a QKD protocol. The method described below can be
employed to generate a continuous stream of key material. This may be
of interest for various practical applications, such as the encryption
of a continuous stream of data.

\paragraph{Description of the Scheme.}

\begin{figure}

\begin{center}
\begin{picture}(375,45)(15, -10)

  \prl{10}{32}{$S^0$}
  \arrowr{15}{22.5}{15}
  
  \arrowr{89.5}{-10}{130}
  \pr{285}{-15}{{\small continuous key stream}}
  \arrowr{240}{-10}{135}

  \system{30}{10}{35}{25}{$\cP_1$}

  \prl{70}{32}{$S^1$}
  \arrowr{65}{22.5}{22}
  \system{87}{20}{5}{5}{}
  \arrowr{92}{22.5}{33}
  \arrowd{89.5}{-10}{30}
  \prl{93}{30}{{\small $\ell_1$ bits}}
  \prl{92}{3}{{\small $\ell$ bits}}

  \system{125}{10}{35}{25}{$\cP_2$}

  \prl{165}{32}{$S^2$}
  \arrowr{160}{22.5}{22}
  \system{182}{20}{5}{5}{}
  \arrowr{187}{22.5}{33}
  \arrowd{184.5}{-10}{30}
  \prl{188}{30}{{\small $\ell_2$ bits}}
  \prl{187}{3}{{\small $\ell$ bits}}

  \pr{230}{22.5}{$\cdots$}  
  \pr{230}{-10}{$\cdots$} 

  \arrowr{242}{22.5}{38}
  \prl{239}{30}{{\small $\ell_{i-1}$ bits}}

  \system{280}{10}{35}{25}{$\cP_i$}

  \prl{320}{32}{$S^i$}
  \arrowr{315}{22.5}{22}
  \system{337}{20}{5}{5}{}
  \arrowr{342}{22.5}{33}
  \arrowd{339.5}{-10}{30}
  \prl{344}{30}{{\small $\ell_i$ bits}}
  \prl{342}{3}{{\small $\ell$ bits}}

  \pr{385}{22.5}{$\cdots$}
  \pr{385}{-10}{$\cdots$}

\end{picture}
\end{center}

\caption{\label{fig:composition} {\bf Generation of a continuous key
    stream by sequential composition of rounds of a QKD protocol.} The
  scheme starts with an initial key pair $S^0 = (S_A^0, S_B^0)$. In
  each round~$i$, the QKD protocol $\cP_i$ generates a fresh pair $S^i
  = (S_A^i, S_B^i)$ of keys of length $\ell + \ell_i$, using
  $\ell_{i-1}$ bits of existing key material for
  authentication. $\ell$ bits of the fresh key are added to the key
  stream, whereas $\ell_i$ bits are passed to the next round for
  authentication.}
\end{figure}
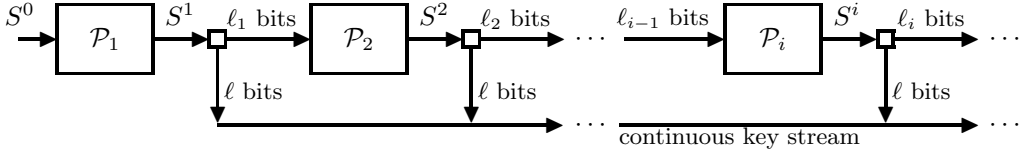

We are looking at the (realistic) situation where the communication
channels connecting Alice and Bob may be completely insecure, so that
not even authenticity is guaranteed.  Instead, we assume that Alice
and Bob hold an initial key pair $(S_A^0, S_B^0)$ of length $\ell_0$
which is $\eps_0$-secure. They then repeat the following for any $i
\in \mathbb{N}$ (see Fig.~\ref{fig:composition}). A QKD protocol
$\cP_i$ is invoked, which uses the first $\ell_{i-1}$ bits of the key
pair $(S_A^{i-1}, S_B^{i-1})$ for authentication. The protocol
generates a new (longer) key pair $(S_A^i, S_B^i)$ of length $\ell_i +
\ell$, of which the first $\ell_i$ bits are stored for use in the next
round, while the last $\ell$ bits form part of the output stream.

\paragraph{Security Analysis.}

In the following, we are going to analyze the security of the key
stream. Because of composability, this is conceptually very easy|we
simply need to add up the security parameters. If the protocol $\cP_i$
executed in each round $i$ is $\eps_i$-secure then the security $\eps$
of the final stream is always bounded by
\begin{align} \label{eq:sum}
  \eps \leq \sum_{i=0}^\infty \eps_i \ .
\end{align}

In order to get a reasonable value for $\eps$, we need to make sure
that the parameters $\eps_i$ are sufficiently small. However, making
$\eps_i$ small generally comes at the cost of increasing the
communication complexity of the protocol as well as the length
$\ell_{i-1}$ of the initial key used for authentication.  As a rough
estimate of the performance of a typical QKD protocol, we use here a
bound of the form
\begin{align} \label{eq:protparam}
  \eps_i \leq e^{-\gamma (\rho n_i - \ell_i - \ell)} + e^{- \nu \ell_{i-1} + \log n_i}
\end{align}
where $n_i$ denotes the number of quantum signals exchanged during the
protocol and where $\gamma$, $\rho$, and $\nu$ are positive
constants.\footnote{Values of $\rho = 10^{-2}$ and $\gamma = \nu =
  10^{-3}$ may be realistic for textbook protocols such as BB84 with
  single photons. We refer to~\cite{ScaRen08,CaiSca08} for a more
  detailed numerical analysis of the performance of QKD protocols.}
The first term corresponds to the security of the protocol if used
with an authentic classical channel. Note that the exponent critically
depends on the length $\ell_i + \ell$ of the key that is
generated. The second term is due to the imperfectness of the
authentication scheme.

To make sure that~\eqref{eq:sum} converges, it is necessary to
increase the number $n_i$ of exchanged signals in each round of the
protocol. For the purpose of illustration,\footnote{The example should
  be understood as a proof of principle. We have not attempted to
  optimize parameters.}  we set
\begin{align*}
  n_i := n + c i \quad \text{and} \quad \ell_i := \ell + c \rho i/2
\end{align*}
for some constants $n \in \mathbb{N}$ and $c > 0$. Inserting this
into~\eqref{eq:protparam} results in a bound on $\eps_i$ such that the
sum over $i$ is a geometric series. Hence, by appropriately choosing
the constants $\ell$, $n$, and $c$, the security parameter $\eps$ of
the key stream can be made arbitrarily small.

\subsection{An Explicit Attack Exploiting
  Non-Composability} \label{sec:attack}

The necessity of composable security definitions has only been
realized recently. In fact, most of the original security proofs
proposed in the literature were relative to a security criterion that
is not composable. The main purpose of this section is to illustrate
what can go wrong if such a non-composable security definition is
used.

\paragraph{Measuring Secrecy.}

As we have seen in Section~\ref{sec:security}, the correctness and the
robustness property are rather unproblematic. In particular, both of
them can be expressed as the condition that certain probabilities are
small (cf.~\eqref{eq:correctness} and~\eqref{eq:robustness}). This is
different for the secrecy property. Intuitively, a key $S_A$ is secret
if an adversary has only \emph{little information} about it, in the
sense of~\eqref{eq:seccrit}. There are, however, a variety of
alternative information measures, and this is indeed the source of the
problem we are going to describe now.

One such information measure is the \emph{accessible information},
denoted ${\Iacc(\cdot:\cdot)}$. It is particularly suitable to
quantify the information a \emph{quantum} system (in our case the
system $E$ held by the adversary) gives about a \emph{classical} value
(the key $S_A$). The accessible information is defined in terms of the
\emph{Shannon mutual information}, ${I(\cdot : \cdot)}$,
\begin{align*}
  \Iacc(S_A : E) := \max_{Z} I(S_A : Z) \ ,
\end{align*}
where the maximum is taken over all random variables $Z$ that can be
obtained by measuring the quantum system $E$.

Recall that, according to Lemma~\ref{lem:secrecy}, the key $S_A$
generated by a QKD protocol is $\eps$-secret if and only if
\begin{align} \label{eq:secrecysimple}
  \frac{1}{2} \bigl\| \rho_{S_A E} - \sum_{s} \frac{1}{|\cS|} \proj{s} \otimes \rho'_E \bigr\|_1 \leq \eps
\end{align}
holds for some $\rho'_E$. (We assume here for simplicity that the
protocol always outputs a key, i.e., $p_{\perp} = 0$.) Since a
measurement cannot increase the trace distance, this immediately gives
a bound on the distance between the joint distribution $P_{S_A Z}$ of
the key $S_A$ and the outcome $Z$ of any measurement applied to $E$,
and a distribution of the form $P_S \times P'_Z$ where $P_S$ denotes a
uniform distribution over the key space,
\begin{align} \label{eq:secrecymeas}
  \frac{1}{2} \bigl\| P_{S_A Z} - P_U \times P'_Z \|_1 \leq \eps \ .
\end{align}
For small values of $\eps$, Fano's inequality implies that $I(S:Z)$
and, hence, the accessible information $\Iacc(S_A : E)$, is small,
too.\footnote{More precisely, \eqref{eq:secrecysimple} implies
  $\Iacc(S_A:E) \leq 2 n \eps + 4 h(\eps)$ where $n$ is the key length
  and $h$ is the binary entropy. (Since $\eps$ is usually chosen
  exponentially small in $n$, the same is true for the term $2 n
  \eps$.)}  In other words, the secrecy
criterion~\eqref{eq:secrecysimple} is at least as strong as a
criterion based on the accessible information.

The converse, however, is not true. To illustrate this, we construct
an explicit example quantum state $\rho_{S_A E}$ for which the
accessible information is (arbitrarily) small, whereas the key $S_A$
is insecure when being used for one-time pad encryption. The state
$\rho_{S_A E}$ thus necessarily violates the (composable) secrecy
criterion~\eqref{eq:secrecysimple}. From this, we conclude that small
accessible information does not imply secrecy in the sense of
Definition~\ref{def:securityq}.

\paragraph{Construction of the Example.}

Our example consists of a uniformly distributed $(n+1)$-bit key $S_A =
(S_1, \ldots, S_{n+1})$ and an $n$-qubit system $E$.  Furthermore, we
consider an $n$-tuple of bits $R = (R_1, \ldots, R_n)$ whose sum
modulo $2$ equals $S_{n+1}$,
\begin{align} \label{eq:Rcond}
  R_1 \oplus \cdots \oplus R_n = S_{n+1} \ ,
\end{align}
but which are otherwise completely random. Then, for any fixed $S_A =
s = (s_1, \ldots, s_n, s_{n+1})$ and $R = r = (r_1, \ldots, r_n)$
satisfying~\eqref{eq:Rcond}, we define the state $\ket{\phi^{s,r}}$ of
$E$ by
\begin{align*}
  \ket{\phi^{s,r}} := \ket{r_1}_{s_1} \otimes \cdots \otimes \ket{r_n}_{s_n} \ ,
\end{align*}
where $\ket{r_i}_{s_i}$, for any $i = 1, \ldots, n$, denotes the state
of a qubit encoding the classical bit $r_i$ in either some specified
standard basis $\{\ket{0}, \ket{1}\}$ (if $s_i = 0$) or the
corresponding diagonal basis (if $s_i = 1$), i.e.,
\begin{align*}
\begin{tabular}{p{5cm}p{5cm}}
{\begin{align*}
  \ket{0}_0 & = \ket{0} \\
  \ket{0}_1 & = \sqrt{\frac{1}{2}}\bigl(\ket{0} + \ket{1}\bigr) 
\end{align*}} &
{\begin{align*}
  \ket{1}_0 & = \ket{1} \\
  \ket{1}_1 & = \sqrt{\frac{1}{2}}\bigl(\ket{0} - \ket{1}\bigr) \ .
\end{align*}}
\end{tabular}
\end{align*}
In particular, the density operator $\rho^s_E$ describing the state of
$E$ conditioned on $S_A = s$ (but randomized over $R$) is given by
\begin{align*}
  \rho^s_E = 2^{-(n-1)} \hspace{-15pt} \sum_{\substack{(r_1, \ldots, r_n) \\ r_1 \oplus \cdots \oplus r_n = s_{n+1}}} \proj{\phi^{s,r}} \ .
\end{align*}

We now move on to the proof of the claims made above. First, we show
that the accessible information $\Iacc(S_A : E)$ is small. This
implies that~\eqref{eq:secrecymeas} holds for some small $\eps$ (see,
e.g., Lemma~12.6.1 of~\cite{CovTho91}). Second, we describe an attack
against a scheme where the key $S_A$ is used for one-time-pad
encryption. The attack allows the adversary to learn one bit of the
message with certainty. This, in particular, implies that the
(composable) secrecy criterion~\eqref{eq:secrecysimple} cannot hold
for any non-trivial value of $\eps$.

\paragraph{Small Accessible Information.} 

We do not attempt here to give a rigorous proof of the above claim but
rather describe the intuition for it. For the details of the argument
we refer to~\cite{KRBM07}.

In order to prove that $\Iacc(S_A : E)$ is small, we need to argue
that any outcome $Z$ of a measurement applied to $E$ has only
negligible correlation with $S_A$.  To simplify this task, we split
$S_A = (S_1, \ldots, S_{n+1})$ into two parts and make use of the
chain rule for the mutual information,
\begin{align*}
  I(S_A : Z) = I(S_1 \cdots S_n : Z) + I(S_{n+1} : Z | S_1 \cdots S_n) \ .
\end{align*}
Note that the state of each qubit of $E$ is an encoding of a random
bit $R_i$, where only the basis depends on $S_i$. The overall state of
$E$ conditioned on $(S_1, \ldots, S_n)$ is thus fully mixed and,
hence, independent of the value of $(S_1, \ldots, S_n)$. This
immediately implies $I(S_1 \cdots S_n : Z) = 0$ and it thus remains to
be shown that $I(S_{n+1} : Z | S_1 \cdots S_n)$ is small. 

For this, let us first assume that the measurement giving $Z$ consists
of $n$ independent measurements applied to the individual qubits of
$E$. Each of them would then result in an estimate for the value of a
bit $R_i$, for $i = 1, \ldots, n$. However, since each bit $R_i$ is
encoded in a random basis determined by $S_i$, and since the bit $S_i$
is unknown at the time of the measurement, the maximum probability $p$
of obtaining the correct outcome $R_i$ is bounded away from $1$, i.e.,
$p < 1$.

Now, recall that the key bit $S_{n+1}$ is equal to the sum modulo $2$
of the random bits $R_1, \ldots R_n$. Hence, using the measurement
strategy described above, the correct value of $S_{n+1}$ can only be
obtained if all the individual measurements are successful. The
probability that this happens can be shown to be exponentially small
$n$.\footnote{More precisely, given $Z$, the probability of correctly
  guessing $S_{n+1}$ is not larger than the probability of guessing an
  independent random bit, except with probability exponentially small
  in $n$.} We thus conclude that the correlation between the key bit
$S_{n+1}$ and the measurement outcome $Z$ is small.

This argument can be generalized to arbitrary measurement
strategies~\cite{KRBM07}. It turns out that the above individual
strategy is essentially optimal, i.e., $I(S_{n+1} : Z | S_1 \cdots
S_n)$ is small for any measurement. In fact, a quantitative
analysis\footnote{For technical reasons, the argument of~\cite{KRBM07}
  is based on an extended construction where the bits $R_i$ are
  encoded with respect to three (rather than two) different mutually
  unbiased bases.} (for a slightly modified example) gives $I(S_A, Z)
< 2^{-\frac{n-2}{6}}$ and, hence, $\Iacc(S_A : E) \leq
2^{-\frac{n-2}{6}}$.

\paragraph{The Attack.}

Let us now have a look at what happens if we use the key $S_A = (S_1,
\ldots, S_{n+1})$ for one-time-pad encryption. By definition, for any
message $M = (M_1, \ldots, M_{n+1})$, the ciphertext $C = (C_1,
\ldots, C_{n+1})$ is given by $C_i = M_i \oplus S_i$. In the
following, we assume that the adversary has full access to $C$.

To understand the relevance of the example, it is important to realize
that we can, in general, not assume that the message $M$ is uniformly
distributed.\footnote{It is possible to design encryption schemes
  whose security is based on the additional assumption that the
  distribution of the messages is highly random from the adversary's
  point of view~\cite{RusWan02} (this is also known as \emph{entropic
    security}). Interestingly, these schemes only require a short
  key.} To the contrary, almost any realistic message will consist of
biased bits or bits that are (partially) known to an adversary. In
fact, the history of cryptography is full of examples where prior
knowledge about the structure of the messages has been exploited for
attacks. For our specific attack, we consider the extreme case where
the adversary already knows the first $n$ message bits $(M_1, \ldots,
M_n)$ but tries to get information about the bit $M_{n+1}$. (For
example, the first $n$ bits may contain standardized header
information while the actual message starts with the $(n+1)$th bit.

Given the first $n$ bits of both the message and the ciphertext, the
adversary can obviously determine the first $n$ key bits $S_1, \ldots,
S_n$ by $S_i = M_i \oplus C_i$. This by itself would not be
problematic because, after all, the very nature of a one-time-pad is
that it is only used once. However, the adversary may now use her
knowledge of $S_1, \ldots, S_n$ to extract further information from
the quantum system $E$. More precisely, because by construction the
bits $S_1, \ldots, S_n$ determine the basis in which the values $R_i$
are encoded in $E$, the adversary can apply a measurement which
produces the outcomes $R_1, \ldots, R_n$. From this, she may determine
the $(n+1)$th key bit $S_{n+1} = R_1 \oplus \cdots \oplus R_n$ and, in
particular, the message bit $M_{n+1} = S_{n+1} \oplus C_{n+1}$ with
certainty.

\paragraph{Discussion.}

Our example shows that the accessible information is an inappropriate
measure for quantifying secrecy: Even tough the accessible information
$\Iacc(S_A, E)$ that an adversary has on the key $S_A$ is small, the
key $S_A$ cannot safely be used for tasks such as one-time-pad
encryption.

The example also answers a question raised by Ben-Or \emph{et al.}
in~\cite{BHLMO05}. They have shown that a QKD protocol which generates
an $n$-bit key $S_A$ is $\eps$-secure whenever
\begin{align*}
  \Iacc(S_A : E) \leq 2^{-(n+2)} \eps^2 \ .
\end{align*}
An immediate implication of our argument above is that this result is
essentially tight.  In other words, in order to get (composable)
security from a bound on $\Iacc(S_A : E)$, this bound must be
exponentially small in the key size.  Unfortunately, however, this
criterion is not met by most known security proofs that refer to the
accessible information (see~\cite{KRBM07} for references).

In order to prove security of a given QKD scheme, it is thus more
advisable to directly derive a bound on the trace distance
in~\eqref{eq:secrecysimple} (rather than on the accessible
information). Such a bound can in principle be obtained by a
modification of the well-known argument by Shor and
Preskill~\cite{ShoPre00}, which however only applies to specific types
of protocols. A more generic approach is to use the fact that privacy
amplification based on suitably chosen hash functions (e.g.,
two-universal hashing) directly produces keys that
satisfy~\eqref{eq:secrecysimple}, provided the input to the hash
function (the \emph{raw key}) has sufficiently high
entropy~\cite{RenKoe05} (see~\cite{DPVR09,TSSR10} for specific
examples of such hash functions).

\section{Composability of General Secure Applications} \label{sec:general}

In the following sections, which constitute the second part of the
article, we consider security definitions for general cryptographic
tasks and the problem of composing secure protocols to complex
security applications.

We will describe a quantum model of
security~\cite{UnruhPreprint04,BenOrMayers04,Unruh10} which gives
strong composability guarantees. The composition theorem (see
Subsection~\ref{subsec:comptheo}) states that a protocol secure in
this model can be used in an arbitrary application without lowering
the overall security. Furthermore an arbitrary number of protocols
proven secure in this model can be used concurrently and remain secure
in the model. We will have to neglect many details
(already~\cite{Canetti01} has 128 pages and describes the classical
case). Our treatment will be on a more intuitive and abstract
level. For details please
see~\cite{UnruhPreprint04,BenOrMayers04,Unruh10}.

One could argue that this topic need not be discussed in an article about quantum cryptography as the most important building blocks of general applications, i.e. protocols like {\em coin flipping}, {\em bit commitment}, or {\em oblivious transfer}, can in quantum cryptography not be achieved with unconditional security~\cite{Ambainis04,Mayers97,LoChau96}. However, there still are enough interesting applications for quantum cryptography. Even if some tasks are impossible to achieve in principle it is possible to achieve them relative to security assumptions which are independent of the computational assumptions of classical cryptography~\cite{Salvail98,Damgaard05}. Furthermore, many of the assumptions possible, like the adversary being able to store only a limited amount of qubits or the adversary being unable to maintain coherency for large quantum states are very reasonable.

In addition a quantum model of security is not only useful to analyze
or prove the security of quantum protocols, but it can also be used to
investigate the security of classical protocols against quantum
adversaries. It was in the context of composability that the question
was answered if quantum attacks on classical protocols give more power
to the adversary than a mere speed up of computations~\cite{Unruh10}
(see Subsection~\ref{subsec:informationtheoretical}).

\section{Defining Security} \label{sec:problem}

Key exchange and secure message transmission is one of the most important prerequisites of general security applications, however, general applications can require further security properties. As examples consider secure authentication, digital signatures, online banking, or remote voting. One of the big differences of such applications to key exchange is that the protocols participants are mutually mistrusting. \emph{Secure function evaluation}~\cite{Yao82,Goldreich04Foundations} is a generalization of such cryptographic applications: In a secure function evaluation a set of players $P_1,\dots,P_n$ wishes to evaluate a function $f$ on inputs $x_1,\dots,x_n$ they hold respectively such that corrupted players cannot change the outcome of the computation (other than choosing a different input) and corrupted players do not learn more about the input of honest players than can be derived from their own input and the output of the function evaluation. These two properties of secure function evaluation are called \emph{correctness} and \emph{privacy}. However, it turned out that these two properties alone do not cover what one intuitively requires from a secure computation. Additional properties were added, like the \emph{independence of inputs} which demands that it should not be possible for a corrupted player to choose his own input dependent on the secret inputs of honest parties. It is easy to see that the property of independence of inputs is not logically implied by privacy or correctness if one does not demand that each protocol participant \emph{knows} its input from the start. There are more security properties which are not implied by privacy and correctness: \emph{robustness} requires that no corrupted player may abort the protocol, \emph{fairness} demands that even if an abort cannot be prevented it should not be possible for the adversary to learn more about the result of the computation than the honest players, and \emph{zero knowledge} is the property that a real protocol transcript could also have been generated by a single machine without knowledge of any secret involved in the protocol. Defining security via a list of security properties became known as the \emph{list approach}, however, researchers got the impression that one might never know if the list of security properties is complete.

\subsection{The Simulation Paradigm}\label{sec:simulation}

A new security definition was needed. It should be convincing and (as general applications are to be considered) independent of the specific goals the attacker might have. The first step towards this new definition was the discovery of zero knowledge proofs~\cite{Goldwasser85} where the \emph{simulation paradigm} was introduced.

Instead of considering different security properties the new notion was based on indistinguishability. 
Intuitively speaking, a real protocol is compared to an ideal protocol where a trusted party collected the inputs from the protocol participants, computes the output and distributes the output to the participants. If the real protocol and the ideal protocol have an indistinguishable input output behavior the real protocol is said to be \emph{at least as secure} as the ideal protocol. Such a definition of security defines security of a real protocol relative to an idealization. The level of security reached thus also depends on the specification of the ideal protocol.

In the case of quantum key distribution we have already seen a
security definition which compares a real key exchange with an ideal
situation, however, unlike to the general case it was possible to
reduce this security notion to the fulfillment of separate security
properties (see Section~\ref{sec:security}).

In the real model the protocol is attacked by a real attacker which may corrupt protocol participants, pools all their data, and lets the corrupted participants deviate from the protocol in an arbitrary way. In the ideal protocol there is a an ideal attacker (also called \emph{simulator}) which must be able to provide an output indistinguishable from the output of the real attacker while having access only to the inputs and outputs of the corrupted players. As the ideal attacker does not learn any real protocol messages or secrets which cannot be derived from the input and output of the corrupted players the indistinguishability guarantees that the real protocol does not leak any secrets to the real attacker.

However, there are certain "attacks" which cannot be prevented, e.g. an adversary could replace his input by a different value. These inevitable attacks are not considered to violate the security and hence we must be able to model these attacks in the ideal protocol as well. These inevitable attacks will be carried out by the simulator, too. The ideal attacker may corrupt protocol participants in the ideal model, but all the ideal attacker can do is to replace local inputs or to replace local outputs. If the real attacker may corrupt more than a minority of the protocol participants then the attacker can always abort the computation and we have to give this ability to the ideal adversary as well.

Stating the exact definition here goes beyond the scope of this article (it can be found in~\cite{Goldreich04Foundations}), especially because this notion of security does not yet allow for composition as we will illustrate below.

Note that this definition of security requires the ideal attacker (simulator) to provide his output only after termination of the protocol, i.e., in retrospect and thus with the benefit of hindsight. This gives a certain "advantage" to the ideal attacker without which a simulation would become impossible in most cases. The ability to provide a simulation of a real protocol without any advantage over a real attacker would in many cases imply the complete insecurity of the real protocol as the real attacker could use the program of the simulator to cheat in the real execution of the protocol. What is important in this context is that this advantage of the simulator should not invalidate the "idealness" of the ideal model.

This simulation in retrospect does not violate the "idealness",
because the result of an ideal protocol is not altered by this (the
protocol remains correct) and no secrets of honest participants are
leaked. However, as we will see in Subsection~\ref{subsection:types},
this ability of simulating in retrospect does not play well with
composition or with protocols which accept inputs not only at the
start, but also at later times (protocols realizing so called
\emph{reactive functionalities} which are a generalization of secure
function evaluation).

\subsection{A Motivating Example: Secure Composition as a Problem}

Below we will give two examples illustrating what can happen when protocols are composed. The first is a classical example from classical cryptography where a message from one subprotocol of a larger application is fed into another subprotocol and the overall application becomes insecure. The second example shows that quantum information can be used in different subprotocols such that entanglement spans over different subprotocols.

\subsubsection{Malleability|a Classical Example}

A very simple example of this kind is an (simplified) auction
protocol. We assume a trusted auctioneer in possession of a RSA public
key $(n, e)$. For an auction the auctioneer accepts bids which are
encrypted with his public key. After receiving all the bids the
auctioneer decrypts the cipher texts with his secret key $d$ and
publishes the highest bid together with the winner of the auction. The
RSA encryption keeps eavesdroppers from learning bids of
competitors. This seems to imply that the bids of the dishonest
participants must be chosen independently of the bids of the honest
participants. However, astonishingly this is not necessarily the case:
Given an honest Alice, a dishonest Bob and let all encryptions be done
by "textbook RSA\footnote{This refers to the originally published
  version of RSA where a ciphertext $c$ for a message $m$ is
  deterministically computed via $c = m^e \mod n$ and decryption is
  done via $m = c^d \mod n$.}". If now Alice bids the amount $m$ then she
sends $c = m^e \mod n$ to the auctioneer. Bob can,
after learning this ciphertext $c$ compute $2^e*c \mod n$ which equals
an encryption of $2*m$ with the public key $(n,e)$.

So without knowing the amount of Alice's bid Bob is able to compute a ciphertext which encrypts a higher bid and so he will win the auction. This security weakness is called \emph{malleability}~\cite{Dolev00} and it is not per se a weakness of textbook RSA, but becomes a problem when textbook RSA is used in certain larger applications.

\subsubsection{Quantum Superpositions can Span over several Subprotocols}

Quantum bit commitment, i.e. the cryptographic equivalent to a sealed
envelope, has been shown to be impossible with unconditional
security. However, it is tempting to try to circumvent this
impossibility theorem of Mayers~\cite{Mayers97} and
Lo/Chau~\cite{LoChau96} by a clever composition of possible quantum
protocols. One could try to build up a secure bit commitment from
weaker primitives like \emph{cheat sensitive
  commitments}~\cite{Hardy04}. However, the impossibility theorem
rules this out and therefore shows that composing quantum protocols
can be counter intuitive. One cannot treat the subprotocols as being
"atomic" and quantum superpositions being limited to occur only within
the subprotocols. It is possible to keep all quantum information in
the different subprotocols in one large superposition and the attack
of Mayers and Lo/Chau does exactly that.

\subsection{Types of Protocol Composition}\label{subsection:types}

Two kinds of protocol composition can be distinguished: 

\emph{Simple Composition} for which an example was given in the
previous subsection. In simple composition a single instance of a
cryptographic primitive is replaced by a real subprotocol.  Now
messages from the surrounding protocol which may depend on secrets of
uncorrupted parties can be injected into the subprotocol or vice
versa: a corrupted player can use messages from within a subprotocol
outside of this subprotocol. This access to protocol messages which
may depend on secrets of uncorrupted parties give an enormous strength
to the adversary not present in stand alone models of security. In the
quantum world it is additionally possible to entangle messages used in
different protocols.

In the case of \emph{Concurrent Composition} many instances of the
same protocol with correlated inputs are run concurrently. Apart from
the problems of simple composition, that messages from one protocol
could be fed into another~\cite{MuellerQuade07}, an additional problem
occurs if one allows more than a constant number of protocol instances
to be run concurrently. Even though each single instance of the
protocol is secure in the sense of simulatability it could be that the
multiple rounds of the different protocol instances are interleaved in
a way that messages in one instance of the protocol affect messages in
other protocols and no polynomial time simulation strategy to obtain a
consistent simulation for all protocols is known.

So in a notion of security allowing for secure composition the simulator should work even if the protocol is run in an arbitrary application context. This implies that the simulation cannot be done in retrospect as the real adversary could feed information into surrounding protocols at any time. This requirement of a \emph{straight line simulator} is very strict, however, according to \cite{Lindell03} it is close to the minimal requirement if one wants to combine the requirements of stand alone simulatability and the notion of security being preserved if run in arbitrary applications.

\section{The Universal Composability Framework}\label{sec:uc}

The basic idea of the \emph{Universal Composability (UC) framework}
and why this notion of security allows for secure composition is that
the \emph{stand alone} simulatability definition of security
from~\cite{Goldreich04Foundations} is enriched by an additional
machine, an \emph{environment machine} which interacts with the
protocol and the attacker while it can emulate arbitrary surrounding
protocols.\footnote{In Section~\ref{sec:comp} this environment was
  only implicit, because the interaction with other protocols is
  simpler than in the general case: key distribution has no input and
  guarantees no security if one of the parties is corrupted.}
Starting from this classical universal composability
framework~\cite{Canetti01} and independently discovered concept of
reactive simulatability\cite{Pfitzmann01,Backes07} two quantum models
of security were defined in~\cite{UnruhPreprint04,BenOrMayers04}. Both
models follow the same motivation, but differ in details which are not
of importance in this overview.

The model of~\cite{UnruhPreprint04} is described in three steps. First
the machines and their network is defined, next the behavior of the
machines is defined according to their roles in a protocol, then the
security definition is given based on the indistinguishability of two
protocols (the real and the ideal protocol). In our overview many
details have to be omitted. For details
consult~\cite{UnruhPreprint04,Unruh10}.

\paragraph{Machines and Networks.} Quantum machines have internal
states which may be quantum and the state transition operator is a
trace preserving superoperator on the Hilbert space spanned by the
tensor product of the possible internal states, the possible inputs
and the possible outputs of the machine. The machines are connected by
an \emph{asynchronous quantum network}, i.e., (quantum) messages
between machines may be blocked or delayed. Only one machine may be
active at any time and the scheduling is \emph{message driven}, i.e.,
a machine sending away a message is switching to a waiting state while
the receiving machine is activated\footnote{One distinguished machine,
  called \emph{master scheduler}, will be invoked if this rule does
  not apply.}.

The scheduling is classical, i.e., machines are not active and inactive in superposition nor are messages sent and not sent in superposition. This makes the model usable, but it excludes the possibility of certain protocols detecting a traffic analysis~\cite{MuellerQuade03,Steinwandt01}.

\paragraph{Protocol, Adversary, and Environment.} Apart from the
protocol participants which are specified by the protocol there are
two more machines taking part in the protocol execution. The
\emph{adversary} $\mathcal{A}$ (or $\mathcal{S}$ in the ideal model)
is the machine coordinating all corrupted participants analogous to
the stand-alone model in Section~\ref{sec:simulation}. The
\emph{environment machine} $\mathcal{Z}$ chooses the
inputs\footnote{In case of a reactive functionality inputs can also
  depend on previous outputs or on protocol messages.}, sees the
output, and may communicate with the adversary at any time. The
environment machine can emulate arbitrary surrounding protocols and
can hence detect vulnerabilities which would result from protocol
composition.

\paragraph{The Security Definition.} We demand the environment machine
to produce a classical output and we say that a protocol $\pi$
implements an ideal protocol $\mathcal{F}$ with \emph{perfect
  security} if for every adversary $\mathcal{A}$ there exists an ideal
adversary $\mathcal{S}$ such that for every environment machine
$\mathcal{Z}$ the distribution of the outputs of $\mathcal{Z}$ when
interacting with $\mathcal{A}$ and $\pi$ equals the distribution of
the outputs of $\mathcal{Z}$ when interacting with $\mathcal{S}$ and
$\mathcal{F}$. A protocol $\pi$ realizes $\mathcal{F}$ with
\emph{statistical security} if the output distribution of
$\mathcal{Z}$ when interacting with $\mathcal{A}$ and $\pi$ is
statistically indistinguishable\footnote{In the case of key
  distribution this amounts to approximate security with $\varepsilon$
  negligible, i.e. asymptotically smaller than any $1/k^n$.} from the
output distribution of $\mathcal{Z}$ when interacting with
$\mathcal{F}$ and $\mathcal{S}$.

Quantum cryptography usually aims at achieving statistical security where the adversary may be limited only by the laws of quantum mechanics. It does, however, make sense to also define \emph{computational security} in the quantum setting, because quantum cryptography can realize tasks with computational security which are believed to be impossible classically\footnote{E.g. realizing oblivious transfer from a one way function~\cite{Yao95,Impagliazzo88}.}.

A machine is said to be quantum polynomial time if it can be invoked only a polynomial number of times in the security parameter $k$ and the input output behavior of the machine can be simulated by a quantum Turing machine in polynomial time in $k$. If now all protocol participants, the adversary and the environment machine are quantum polynomial machines then we say that a protocol $\pi$ realizes $\mathcal{F}$ with \emph{quantum computational security} if  for all $\mathcal{A}$ there exists a $\mathcal{S}$ such that for all $\mathcal{Z}$ the output distribution of $\mathcal{Z}$ when interacting with $\mathcal{A}$ and $\pi$ is indistinguishable in quantum polynomial time from the output distribution of $\mathcal{Z}$ when interacting with $\mathcal{F}$ and $\mathcal{S}$. I.e. if we denote by $out_{\pi,\mathcal{A},\mathcal{Z}}$ the random variable of the output of $\mathcal{Z}$ in the real protocol and by $out_{\mathcal{F},\mathcal{S},\mathcal{Z}}$ the corresponding random variable for the ideal model then we demand that for every quantum polynomial machine $\mathcal{D}$ it holds that $|P(\mathcal{D}(out_{\pi,\mathcal{A},\mathcal{Z}})\rightarrow 1) -P(\mathcal{D}(out_{\mathcal{F},\mathcal{S},\mathcal{Z}})\rightarrow 1)|$ is negligible in the security parameter (where a function $\epsilon$ is called negligible if it is asymptotically smaller than any $1/k^n$ for every constant $n$).

\subsection{The Composition Theorem}\label{subsec:comptheo}

The UC framework provides a very strict notion of security and for a
protocol $\rho$ securely realizing an ideal protocol $\mathcal{F}$ in
the UC framework strong composition guarantees can be obtained. We
denote by $\pi^\mathcal{F}$ that a protocol $\pi$ invokes a protocol
$\mathcal{F}$ as a subprotocol and by $\pi^\rho$ that $\mathcal{F}$
has been replaced by a protocol $\rho$. We write $\pi \geq \rho$ to
denote that the protocol $\pi$ securely realizes $\rho$ in the UC
framework. Now the (simple) composition theorem (see
\cite{UnruhPreprint04,Unruh10}) states that if $\rho \geq \mathcal{F}$
then $\pi^\rho$ securely realizes $\pi^\mathcal{F}$. Especially if
$\pi^\mathcal{F}$ securely realizes a functionality $\mathcal{G}$ then
also $\pi^\rho$ realizes $\mathcal{G}$.

If we denote by $\rho^*$ the concurrent composition of (polynomially
many) instances of $\rho$ and by $\mathcal{F}^*$ the concurrent
composition of (polynomially many) instances of $\mathcal{F}$. Then
the (concurrent) composition theorem guarantees that if $\rho \geq
\mathcal{F}$ it also holds that $\rho^* \geq \mathcal{F}^*$.

Combining simple and concurrent composition we obtain the composition theorem where a larger application $\pi$ may use multiple instances of a subprotocol: Given a protocol $\rho$ which securely realizes a protocol $\mathcal{F}$ in the UC framework, then a protocol $\pi^{\rho^*}$ securely realizes $\pi^{\mathcal{F}^*}$ in the UC framework.

The UC framework is to a certain extent a minimal requirement for the
composition theorem. In the classical case it was shown in
\cite{Lindell03} that a security notion comparable to the UC framework
naturally arises if one demands stand alone simulatability (see
Section~\ref{sec:simulation}) and the existence of a composition
theorem.

\subsection{Information Theoretical Security and Quantum Adversaries}\label{subsec:informationtheoretical}

One very interesting result proven in the quantum universal composability framework regards the security of classical protocols with respect to a quantum adversary. Given a protocol which is proven to be statistically secure against a classic adversary. Does it remain secure under quantum attacks? Is the speed-up of quantum computing the only threat to classical protocols or could a quantum attacker together with a quantum environment use entangled quantum information to break classical protocols?

In \cite{Unruh10} it was shown that whenever a protocol
$\rho$ realizes some ideal protocol $\mathcal{F}$ with respect to
statistical security in the UC framework, then $\rho$ securely
realizes $\mathcal{F}$ in the quantum composability setting.

This result is very useful. Quantum Key Distribution (QKD) is
composable (cf.\ Section~\ref{sec:comp}) and from QKD one can obtain
composable secure communication~\cite{Raub05}. Hence secure channels
based on quantum cryptography can be used instead of idealized secure
channels in many cryptographic settings, such as secure multiparty
computations in presence of an honest majority~\cite{Chaum88}.

\subsection{Impossibility of Bit Commitment}

Additionally to the impossibility of unconditionally secure bit commitment in quantum cryptography~\cite{Mayers97,LoChau96} a new impossibility result is introduced by the UC framework: Without additional security assumptions bit commitment cannot be realized with computational security~\cite{CanettiFischlin01}. This result generalizes to many more cryptographic tasks like coin flipping or oblivious transfer and it also holds in the quantum case.

The reason for this impossibility result is that the simulator may no more act in retrospect and without additional assumptions every simulation strategy for $\mathcal{S}$ could be turned into a cheating strategy for the adversary $\mathcal{A}$ in the real protocol.

The additional assumptions used to allow for a computationally secure
bit commitment can be a trusted authority providing randomness to the
protocol participants before the start of the protocol (the \emph{Common Reference String (CRS)})~\cite{CanettiFischlin01}, a trusted
authority setting up a trusted public key infrastructure, or the
availability of tamper proof hardware. What is worse such \emph{set-up
  assumptions} are needed in quantum cryptography, too.  The
impossibility result of~\cite{CanettiFischlin01} directly carries over
to the quantum case thus in the UC framework quantum cryptographic
protocols cannot even achieve a computationally secure bit commitment
without additional security assumptions.

So for many cryptographic tasks where the protocol participants are
mutually mistrusting one has a trade-off between the strength of the
composability guarantees and the strength of the assumptions needed to
achieve these tasks. For certain applications the threats introduced
by the additional assumptions (e.g. the trusted authorities) weigh
heavier than the threats introduced by improper composition of
protocols and it seems that for this case there is no security notion
which is without a compromise.

As we will see in the next subsection the above impossibility result
also affects the composability of protocols in the bounded quantum
storage model~\cite{Damgaard05}. To allow for simulatable security in
the bounded quantum storage model the memory restrictions have to be
different in the real and in the ideal model, which results in
difficulties when applying the composition theorem multiple times.

\subsection{Composability in the Bounded Quantum Memory Model}\label{sec:boundedquantumstorage}

Even though many interesting cryptographic tasks are not realizable
from scratch these tasks can be realized under very reasonable
security assumptions, e.g. that the adversary is limited in performing
large coherent operations~\cite{Salvail98} or that the adversary has a
quantum memory which is bounded in size~\cite{Damgaard05}. It was
shown that the protocols in the bounded quantum storage model do
compose sequentially~\cite{Wehner08}, however, the protocols as stated
do not allow general composition. With an example we will illustrate
that this seems to be a general problem. To have a useful composition
theorem we need that the \emph{at least as secure as} relation
($\geq$) is transitive, because otherwise we cannot repeatedly apply
the composition theorem in the modular design of a cryptographic
protocol. To be able to conclude from $\pi\geq\rho$ and $\rho\geq F$
that $\pi$ securely realizes $F$ we need that the simulator in the
protocol $\rho$ should be admitted as a real adversary for $\rho$ if
this protocol is to be compared with $F$. In~\cite{Hofheinz03} it is
shown that it is possible to achieve oblivious transfer (and hence bit
commitment) if the real adversary is restricted to have no quantum
memory at all. However, the simulator for this protocol needs quantum
memory for the simulation. So if we restrict the simulator to have no
quantum memory oblivious transfer is not realizable any more and
having different restrictions for the real attacker and the simulator
results in $\geq$ not being transitive. A way around this problem is
to generalize the notion of \emph{at least as secure as} to one that
explicitly involves the memory bound of the adversary as a parameter,
as proposed in~\cite{Unruh10b}.

\section{Conclusions}

This work reviewed composable security in quantum cryptography. In the
first part of the paper the focus was on quantum key distribution
(QKD), the most prominent application of quantum cryptography. We
discussed the requirements that a composable security definition must
fulfill and illustrated the importance of these requirements by an
attack which exploits a typical weakness of a non-composable (but
widely used) definition for secrecy. To show the utility of composable
security, we constructed a scheme to generate a continuous key stream
by sequentially composing rounds of a quantum key distribution
protocol.

The second part of the work took a more general point of view, which is
necessary for the study of security applications involving general
tasks as well as mutually distrustful parties. We explained the
universal composability framework and stated its composition theorem
which gives strong composability guarantees. Of special interest was
the secure composition of quantum protocols into unconditionally
secure classical protocols. This shows that every unconditionally
secure protocol possible in the secure channel model is also possible
with QKD and does not even need a new proof.
 
However, there are open problems left. A drawback of the universal
composability framework is that some tasks become impossible there
without adding new security assumptions. E.g., quantum bit commitment
is impossible in the universal composability framework even with mere
computational security or with respect to an attacker in the bounded
quantum storage model. Hence we observe a trade-off between the strong
guarantees provided by universal composability and the possibility of
using fewer security assumptions. Addressing this trade-off remains an
open problem. A concrete approach may be to consider additional (weak)
setup assumptions, e.g., a Common Reference String as used in the
classical model~\cite{CanettiFischlin01}.

Another open question regards a weakness inherent to most existing
security proofs in quantum cryptography. These proofs typically rely
on a specific model for the hardware the scheme is built on (e.g., the
photon sources and detectors used for optical QKD). Obviously, the
security claims derived for such a model generally only apply to
implementations that strictly match the model. This, however, is
almost never the case in practice.  Indeed, explicit attacks
exploiting the deviation of the implementation from the theoretical
model have been demonstrated recently (see, e.g.,
~\cite{Zhao08,Makarov09}). It would thus be desirable to have a
(composable) framework that allows a more flexible modeling of the
underlying hardware devices.

\paragraph{Acknowledgments}

We would like to thank Gilles Brassard for helpful comments on an
earlier version of the manuscript.


\begin{thebibliography}{10}

\bibitem{Ambainis04}
Andris Ambainis, Harry Buhrmann, Yevgeniy Dodis, and Hein R{\"o}hrig.
\newblock Multiparty quantum coin flipping.
\newblock In {\em 19th Annual IEEE Conference on Computational Complexity (CCC
  2004)}, pages 250--259, 2004. 

\bibitem{Backes07}
Michael Backes, Birgit Pfitzmann, and Michael Waidner.
\newblock The reactive simulatability (rsim) framework for asynchronous
  systems.
\newblock {\em Inf. Comput.}, 205(12):1685--1720, 2007.

\bibitem{BHLMO05}
Michael Ben-Or, Michal Horodecki, Debbie~W. Leung, Dominic Mayers, and Jonathan
  Oppenheim.
\newblock The universal composable security of quantum key distribution.
\newblock In {\em Second Theory of Cryptography Conference {TCC}}, volume 3378
  of {\em LNCS}, pages 386--406, 2005.

\bibitem{BenOrMayers04}
Michael Ben-Or and Dominic Mayers.
\newblock General security definition and composability for quantum \&
  classical protocols.
\newblock arXiv:quant-ph/0409062, 2004.

\bibitem{BenBra84}
Charles~H. Bennett and Gilles Brassard.
\newblock Quantum cryptography: Public-key distribution and coin tossing.
\newblock In {\em Proceedings of IEEE International Conference on Computers,
  Systems and Signal Processing}, pages 175--179, 1984.

\bibitem{BihMor97a}
Eli Biham and Tal Mor.
\newblock Security of quantum cryptography against collective attacks.
\newblock {\em Phys.\ Rev.\ Lett.}, 78(11):2256--2259, 1997.

\bibitem{Blum83}
Manuel Blum.
\newblock Coin flipping by telephone: a protocol for solving impossible
  problems.
\newblock {\em ACM SIGACT News}, 15:23--27, 1983.

\bibitem{CaiSca08}
Raymond Y.~Q. Cai and Valerio Scarani.
\newblock Finite-key analysis for practical implementations of quantum key
  distribution.
\newblock {\em New J. Phys.}, 11:045024, 2009.

\bibitem{Canetti01}
Ran Canetti.
\newblock Universally composable security: A new paradigm for cryptographic
  protocols.
\newblock In {\em 42nd Annual Symposium on Foundations of Computer Science,
  FOCS 2001}, pages 136--145, 2001.

\bibitem{CanettiFischlin01}
Ran Canetti and Marc Fischlin.
\newblock Universally composable commitments.
\newblock In {\em Advances in Cryptology --- CRYPTO 2001}, volume 2139 of {\em
  LNCS}, pages 19--40, 2001.

\bibitem{Chaum88}
David Chaum, Claude Cr{\'e}peau, and Ivan Damg{\aa}rd.
\newblock Multiparty unconditionally secure protocols (extended abstract).
\newblock In {\em Proceedings of the Twentieth Annual ACM Symposium on Theory
  of Computing (STOC)}, pages 11--19, 1988.

\bibitem{CovTho91}
Thomas~M. Cover and Joy~A. Thomas.
\newblock {\em Elements of Information Theory}.
\newblock Wiley Series in Telecommunications. Wiley, New York, 1991.

\bibitem{Damgaard05}
Ivan~B. Damg{\aa}rd, Serge Fehr, Louis Salvail, and Christian Schaffner.
\newblock Cryptography in the bounded quantum-storage model.
\newblock In {\em 46th Annual IEEE Symposium on Foundations of Computer Science
  (FOCS)}, pages 449--458, 2005. 

\bibitem{DPVR09}
Anindya De, Christopher Portmann, Thomas Vidick, and Renato Renner.
\newblock Trevisan's extractor in the presence of quantum side information.
\newblock arXiv:0912.5514, 2009.

\bibitem{Dolev00}
Danny Dolev, Cynthia Dwork, and Moni Naor.
\newblock Nonmalleable cryptography.
\newblock {\em SIAM J. Comput.}, 30(2):391--437, 2000.

\bibitem{Ekert91}
Artur Ekert.
\newblock Quantum cryptography based on {Bell's} theorem.
\newblock {\em Phys.\ Rev.\ Lett.}, 67:661--663, 1991.

\bibitem{Goldreich04Foundations}
Oded Goldreich.
\newblock {\em {Foundations of Cryptography -- II Basic Applications}},
  volume~2.
\newblock Cambridge University Press, 2004.

\bibitem{Goldwasser85}
Shafi Goldwasser, Silvio Micali, and Charles Rackoff.
\newblock The knowledge complexity of interactive proof-systems (extended
  abstract).
\newblock In {\em Proceedings of the Seventeenth Annual ACM Symposium on Theory
  of Computing (STOC)}, pages 291--304, 
  1985. 

\bibitem{Hardy04}
Lucien Hardy and Adrian Kent.
\newblock Cheat sensitive quantum bit commitment.
\newblock {\em Phys. Rev. Lett.}, 92(157901), 2004.

\bibitem{Hofheinz03}
Dennis Hofheinz and J{\"o}rn M{\"u}ller-Quade.
\newblock A paradox of quantum universal composability.
\newblock Poster at the 4th European QIPC Workshop, 2003. Abstract online at
  http://www.quiprocone.org/Oxford/Abstracts.htm, 2003.

\bibitem{Impagliazzo88}
Russell Impagliazzo and Steven Rudich.
\newblock Limits on the provable consequences of one-way permutations.
\newblock In Shafi Goldwasser, editor, {\em Advances in Cryptology --- CRYPTO
  1988}, LNCS, 1988. 

\bibitem{Kao08}
Ming-Yang Kao, editor.
\newblock {\em Encyclopedia of Algorithms}, chapter Quantum key distribution.
\newblock Springer, 2008.

\bibitem{KRBM07}
Robert K\"onig, Renato Renner, Andor Bariska, and Ueli Maurer.
\newblock Small accessible quantum information does not imply security.
\newblock {\em Phys.\ Rev.\ Lett.}, 98:140502, 2007.

\bibitem{Lindell03}
Yehuda Lindell.
\newblock General composition and universal composability in secure multi-party
  computation.
\newblock In {\em 44th Symposium on Foundations of Computer Science (FOCS
  2003)}, 2003. 

\bibitem{LoChau96}
Hoi-Kwong Lo and H.~F. Chau.
\newblock Why quantum bit commitment and ideal quantum coin tossing are
  impossible.
\newblock In {\em PhysComp96: Proceedings of the Fourth Workshop on Physics and
  Computation}, pages 177--187,  1998. 

\bibitem{Makarov09}
Vadim Makarov.
\newblock Controlling passively quenched single photon detectors by bright
  light.
\newblock {\em New J. Phys.}, 11:065003, 2009.

\bibitem{Maurer02}
Ueli Maurer.
\newblock Indistinguishability of random systems.
\newblock In Lars Knudsen, editor, {\em Advances in Cryptology ---
  \mbox{EUROCRYPT 2002}}, volume 2332 of {\em LNCS}, pages 110--132,
  2002.

\bibitem{Mayers97}
Dominic Mayers.
\newblock Unconditionally secure bit commitment is impossible.
\newblock {\em Phys. Rev. Lett.}, 78:3414--3417, 1997.

\bibitem{MuellerQuade03}
J{\"o}rn M{\"u}ller-Quade and Rainer Steinwandt.
\newblock On the problem of authentication in a quantum protocol to detect
  traffic analysis.
\newblock {\em Quantum Information and Computation}, 3(1):48--54, 2003.

\bibitem{MuellerQuade07}
J{\"o}rn M{\"u}ller-Quade and Dominique Unruh.
\newblock Long-term security and universal composability.
\newblock In {\em 4th Theory of Cryptography Conference, TCC 2007}, LNCS, pages
  41--60, 2007.

\bibitem{Pfitzmann01}
Birgit Pfitzmann and Michael Waidner.
\newblock A model for asynchronous reactive systems and its application to
  secure message transmission.
\newblock In {\em Proceedings of the 2001 IEEE Symp. on Security and Privacy}, 2001.

\bibitem{Raub05}
Dominik Raub, Rainer Steinwandt, and J{\"o}rn M{\"u}ller-Quade.
\newblock On the security and composability of the one time pad.
\newblock In {\em 31st Conference on Current Trends in Theory and Practice of
  Computer Science, SOFSEM 2005}, volume 3381 of {\em LNCS}, pages 288--297, 2005.

\bibitem{Renner05}
Renato Renner.
\newblock {\em Security of Quantum Key Distribution}.
\newblock PhD thesis, Swiss Federal Institute of Technology (ETH) Zurich, 2005.
\newblock Electronic version: arXiv:quant-ph/0512258.

\bibitem{Renner07}
Renato Renner.
\newblock Symmetry of large physical systems implies independence of
  subsystems.
\newblock {\em Nature Physics}, 3(9):645--649, 2007.

\bibitem{RenKoe05}
Renato Renner and Robert K\"onig.
\newblock Universally composable privacy amplification against quantum
  adversaries.
\newblock In {\em Second Theory of Cryptography Conference {TCC}}, volume 3378
  of {\em LNCS}, pages 407--425. 2005.

\bibitem{RusWan02}
Alexander Russell and Hong Wang.
\newblock How to fool an unbounded adversary with a short key.
\newblock In {\em Advances in Cryptology --- EUROCRYPT 2002}, volume 2332,
  pages 133--148, 2002.

\bibitem{Salvail98}
Louis Salvail.
\newblock Quantum bit commitment from a physical assumption.
\newblock In {\em Advances in Cryptology --- CRYPTO 1998}, volume 1462 of {\em
  LNCS}, pages 338--353, 1998.

\bibitem{SBCDLP08}
Valerio Scarani, Helle Bechmann-Pasquinucci, Nicolas~J. Cerf, Miloslav Dusek,
  Norbert L\"utkenhaus, and Momtchil Peev.
\newblock The security of practical quantum key distribution.
\newblock {\em Rev. Mod. Phys.}, 81, 2009.

\bibitem{ScaRen08}
Valerio Scarani and Renato Renner.
\newblock Quantum cryptography with finite resources: Unconditional security
  bound for discrete-variable protocols with one-way postprocessing.
\newblock {\em Phys.\ Rev.\ Lett.}, 100:200501, 2008.

\bibitem{ShoPre00}
Peter~W. Shor and John Preskill.
\newblock Simple proof of security of the {BB84} quantum key distribution
  protocol.
\newblock {\em Phys.\ Rev.\ Lett.}, 85:441--444, 2000.

\bibitem{Steinwandt01}
Rainer Steinwandt, Dominik Janzing, and Thomas Beth.
\newblock On using quantum protocols to detect traffic analysis.
\newblock {\em Quantum Information and Computation}, 1(3):62--69, 2001.

\bibitem{TSSR10}
Marco Tomamichel, Christian Schaffner, Adam Smith, and Renato Renner.
\newblock Leftover hashing against quantum side information.
\newblock arXiv:1002.2436, 2010.

\bibitem{UnruhPreprint04}
Dominique Unruh.
\newblock Simulatable security for quantum protocols.
\newblock arXiv:quant-ph/0409125, 2004.

\bibitem{Unruh10b}
Dominique Unruh.
\newblock Concurrent composition in the bounded quantum storage model.
\newblock http://eprint.iacr.org/2010/229, 2010.

\bibitem{Unruh10}
Dominique Unruh.
\newblock Universally composable quantum multi-party computation.
\newblock In {\em Advances in Cryptology --- EUROCRYPT 2010}, volume 6110 of
  {\em LNCS}, pages 486--505, 2010.

\bibitem{WegCar81}
Mark~N. Wegman and J.~Lawrence Carter.
\newblock New hash functions and their use in authentication and set equality.
\newblock {\em Journal of Computer and System Sciences}, 22:265--279, 1981.

\bibitem{Wehner08}
Stephanie Wehner and J{\"u}rg Wullschleger.
\newblock Composable security in the bounded-quantum-storage model.
\newblock In {\em Automata, Languages and Programming, 35th International
  Colloquium, ICALP 2008}, pages 604--615, 2008.

\bibitem{Wiesner83}
Stephen Wiesner.
\newblock Conjugate coding.
\newblock {\em Sigact News}, 15(1):78--88, 1983.

\bibitem{Yao82}
Andrew Chi~Chih Yao.
\newblock Protocols for secure computations (extended abstract).
\newblock In {\em Proceedings of the 14th Annual ACM Symposium on Theory of
  Computing (STOC)}, 1982.

\bibitem{Yao95}
Andrew Chi~Chih Yao.
\newblock Security of quantum protocols against coherent measurements.
\newblock In {\em Proceedings of 26th Annual ACM Symposium on the Theory of
  Computing (STOC)}, pages 67--75, 1995. 

\bibitem{Zhao08}
Yi~Zhao, Chi Hang~Fred Fung, Bing Qi, Christine Chen, and Hoi~Kwong Lo.
\newblock Quantum hacking: Experimental demonstration of time shift attack
  against practical quantum key-distribution systems.
\newblock {\em Phys. Rev. A}, 78:04233, 2008.

\end{thebibliography}

\end{document}